\documentclass[pre,twocolumn]{revtex4-2}

\usepackage{amsmath,amsfonts,amssymb}
\usepackage{bm}
\usepackage{mathrsfs}
\usepackage{graphicx}
\usepackage{hyperref}
\usepackage{braket}
\usepackage{mathtools}

\usepackage[whole]{bxcjkjatype}
\usepackage[normalem]{ulem}
\usepackage[usenames,dvipsnames]{xcolor}
\usepackage{soul}

\hypersetup{
    colorlinks,
    citecolor=blue,
    linkcolor=blue,
    urlcolor=blue
}

\begin{document}
\title{Quantum Otto cycle under strong coupling}
\author{Mao Kaneyasu}
\email{kaneyasu@biom.t.u-tokyo.ac.jp}
\affiliation{Department of Information and Communication Engineering, Graduate
School of Information Science and Technology, The University of Tokyo,
Tokyo 113-8656, Japan}

\author{Yoshihiko Hasegawa}
\email{hasegawa@biom.t.u-tokyo.ac.jp}
\affiliation{Department of Information and Communication Engineering, Graduate
School of Information Science and Technology, The University of Tokyo,
Tokyo 113-8656, Japan}
\date{\today}

\begin{abstract}

Quantum heat engines are often discussed under the weak coupling assumption that the interaction between the system and the reservoirs is negligible.
Although this setup is easier to analyze, this assumption cannot be justified on the quantum scale.
In this study, a quantum Otto cycle model that can be generally applied without the weak coupling assumption is proposed.
We replace the thermalization process in the weak coupling model with a process comprising thermalization and decoupling.
The efficiency of the proposed model is analytically calculated and it indicates that when the contribution of the interaction terms is neglected in the weak interaction limit, it reduces to that of the earlier model.
The sufficient condition for the efficiency of the proposed model not to surpass that of the weak coupling model is that the decoupling processes of our model have a positive cost.
Moreover, the relation between the interaction strength and the efficiency of the proposed model is numerically examined using a simple two-level system.
Furthermore, we show that our model's efficiency can surpass that of the weak coupling model under particular cases.
From analyzing the majorization relation, we also find a design method of the optimal interaction Hamiltonians which are expected to provide the maximum efficiency of the proposed model.
Under these interaction Hamiltonians, the numerical experiment shows that the proposed model achieves higher efficiency than that of its weak coupling counterpart.

\end{abstract}

\maketitle

\section{Introduction}

Constructing and analyzing heat engines is one of the fundamental themes in thermodynamics.
In classical thermodynamics, it is a universal principle, rigorously shown by Carnot, that no heat engine operating between two reservoirs can exceed the efficiency limit $\eta_C = 1 - T_c/T_h$, where $T_c$ and $T_h$ denotes the temperatures of the cold and hot reservoirs, respectively. 
The Carnot limit assumes that heat engines operate at the macroscopic scale, where fluctuations and quantum effects do not come into play. 
Recently, the notion of thermodynamics has been applied to mesoscopic systems, such as protein motors and biochemical clocks, that are described by stochastic processes.
In the mesoscopic regime, thermodynamic quantities, e.g., entropy, work, and heat, become stochastic and the second law of thermodynamics does not necessarily hold due to fluctuations \cite{ritort2008review,seifert2012review}.
Moreover, heat engines have been studied in microscopic systems, where the quantum effects play fundamental roles \cite{quan2007quantum,gardas2015thermodynamic,pena2020otto,ding2018measurement,anka2021measurement,buffoni2019quantum,yi2017single,huang2012effects,rossnagel2014nanoscale,klaers2017squeezed,huang2014quantum,camati2019coherence,denzler2020efficiency,mitchison2019quantum,saryal2021bounds,campisi2015nonequilibrium,kose2019algorithmic,quan2009quantum,das2019measurement,kosloff2017quantum,deffner2018efficiency,lee2021quantumness}.
Quantum extensions of the Carnot and Otto cycles, the most fundamental heat engines in thermodynamics, are summarized in Refs.~\cite{quan2007quantum,gardas2015thermodynamic,pena2020otto}.
Additionally, various heat engines that utilize quantum effects, such as measurement, coherence, and entanglement, have been proposed \cite{ding2018measurement,anka2021measurement,buffoni2019quantum,yi2017single}.
In particular, the achievable efficiency of quantum heat engines has been theoretically proven to possibly exceed the classical efficiency limit \cite{gardas2015thermodynamic,huang2012effects,rossnagel2014nanoscale,klaers2017squeezed,huang2014quantum}.
For instance, the classical Carnot limit is violated in heat engines using squeezed reservoirs \cite{huang2012effects, rossnagel2014nanoscale,klaers2017squeezed}, although this phenomenon does not violate the second law of thermodynamics.
This fact indicates that quantum resources can be used to enhance heat engines.
Currently, research on the experimental realization of quantum heat engines is ongoing \cite{rossnagel2016single,abah2012single,peterson2019experimental,von2019spin,de2019efficiency,klaers2017squeezed} and quantum heat engines have already been implemented in various physical platforms, such as trapped ions and nuclear magnetic resonance (see Ref. \cite{myers2022quantum} for a review).

In many quantum heat engine models, the interaction between the system and the reservoirs is assumed to be weak to ensure that their interaction is negligible.
This approximation facilitates the theoretical analysis, because the thermal equilibrium state can be regarded as the product state of the system and the reservoir.
However, this assumption cannot be justified in systems where quantum behavior appears, i.e., the effects of the interaction cannot be negligible inevitably in the quantum scale.
That is because the ratio of the surface to the volume is large when the volume of the system is very small \cite{strasberg2016nonequilibrium,perarnau2018strong}.
In recent years, theories do not assume weak coupling but consider the contribution of the interaction \cite{rivas2020strong,perarnau2018strong,newman2017performance,newman2020quantum,gelbwaser2015strongly,gallego2014thermal,strasberg2016nonequilibrium,xu2018achieving,katz2016quantum,seifert2016first,carrega2016energy}.
One representative method is the reaction coordinate mapping \cite{nazir2018reaction}.
In this method, a reaction coordinate is introduced to account for the contributions of the interactions between a system and reservoirs. 
After this mapping, the strongly coupled system and reservoirs can be treated as if the reaction coordinate couples to the residual environments weakly.
Despite the method being applicable to arbitrary quantum systems, it has a restriction that the reservoirs, consisting of harmonic oscillators, must be linearly coupled to the system.
To the best of our knowledge, more general treatment applicable to anharmonic or nonlinear baths \cite{makri1998semiclassical,bhadra2016system,xu2018theories} has not yet been proposed.

In this study, we construct a quantum Otto cycle model without making the approximation that the interaction between the system and the reservoirs is negligible.
This model can be applied generally: the cycle is analyzed using density operators throughout to ensure that the model does not specify the details of the system and the reservoirs, except that the decoupling processes are assumed to be realized under the Schr\"odinger equation.
To confirm the consistency, the proposed model is shown to agree with the existing model in the weak coupling limit.
In addition, the sufficient condition is derived for $\eta_\mathrm{str} \leq \eta_\mathrm{weak}$, where $\eta_\mathrm{str}$ and $\eta_\mathrm{weak}$ are the efficiencies of our model (strong coupling) and the existing model (weak coupling), respectively.
This inequality holds for positive costs to decouple the system from the two reservoirs.
In the numerical experiment, the interaction provides a detrimental effect to the strong coupling model.
Furthermore, the efficiency of the strong coupling model is visually demonstrated to be lower than that of the weak coupling model.

Although the relation $\eta_\mathrm{str} \leq \eta_\mathrm{weak}$ is primarily satisfied, we also suggest the possibility that the efficiency of the proposed model can be higher than that of the existing weak coupling counterpart.
This insight is obtained by considering the majorization relation, which is originally a mathematical concept but plays a significant role in thermodynamics, especially in resource theory (see Refs. \cite{marshall1979inequalities,sagawa2022entropy} for review).
Finally, we propose a design method to realize the optimal interaction Hamiltonians, which are expected to provide the maximum efficiency, and the reversal of efficiency $\eta_\mathrm{str} > \eta_\mathrm{weak}$ is numerically demonstrated to be realized under these interaction Hamiltonians.

\section{Weak coupling model}

This section reviews the widely discussed quantum Otto cycle model.
In this model, the interaction between the system and the reservoirs is assumed to be negligible.
We call this the ``weak coupling model'' to clearly distinguish it from the model described in the next section.

Consider a quantum system $S$ and two heat reservoirs $B_h$ and $B_c$, whose inverse temperatures are $\beta_h$ and $\beta_c$, respectively ($\beta_h < \beta_c$).
The Hamiltonian of the total system is expressed as follows:
\begin{equation}
  H_\mathrm{tot} = H_S + H_B^h + H_B^c + H_{SB}^h + H_{SB}^c.
\end{equation}
$H_S$, $H_B^h$, and $H_B^c$ are the self-Hamiltonians of $S$, $B_h$, and $B_c$, respectively, and $H_{SB}^i$ corresponds to the interaction between $S$ and $B_i$ $(i=h,c)$.
By assuming that the two reservoirs are both in Gibbs states, the states of the reservoirs can be expressed as
\begin{equation}
  \rho_B^i = \frac{e^{-\beta_iH_B^i}}{Z_B^i}\;\;(i=h,c),
\end{equation}
where $Z_B^i = \mathrm{Tr}[e^{-\beta_iH_B^i}]$ is the partition function.

In this model, the cycle consists of the following four processes: (A) adiabatic compression, (B) hot isochoric thermalization, (C) adiabatic expansion, and (D) cold isochoric thermalization.
The state change of the system in each process is described, and the transferred heat and the exerted work during these processes are calculated below.

\begin{figure}
  \includegraphics[width=8.5cm]{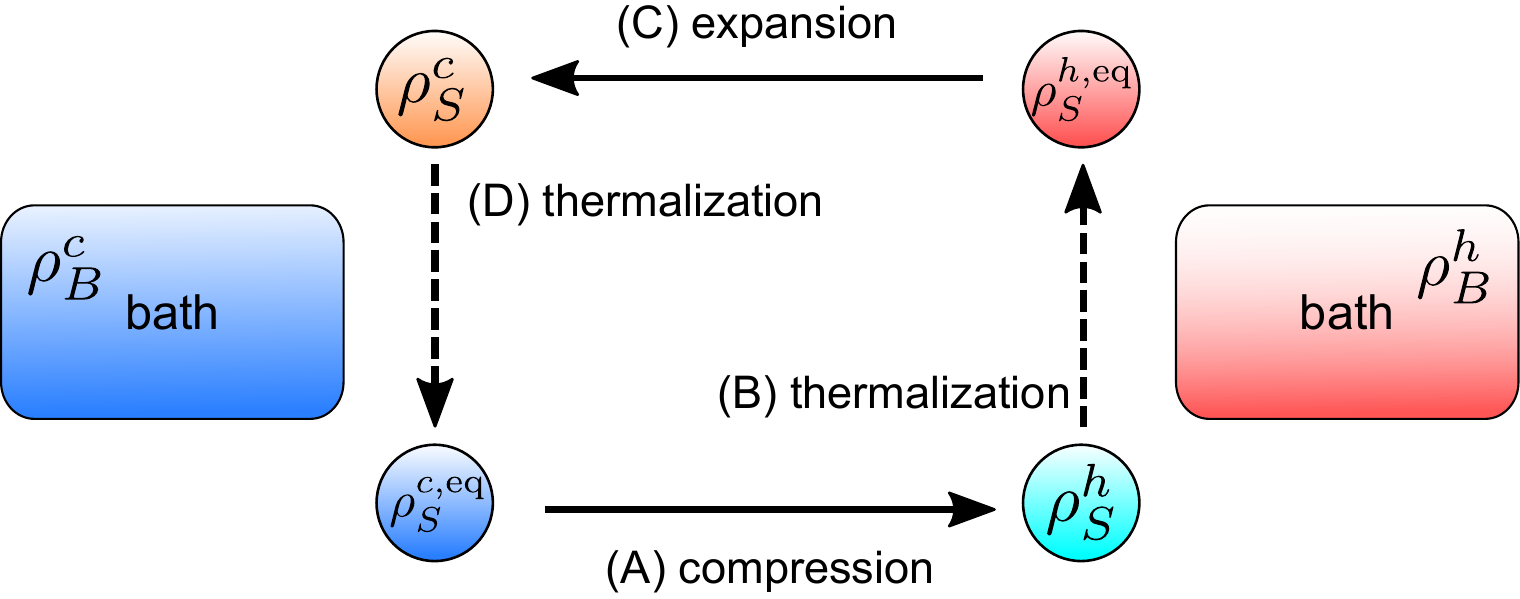}
  \caption{Four processes in the weak coupling model. The state of the two reservoirs is constant and the state of the system transitions as follows. (A) Adiabatic compression: $\rho_S^{c,\mathrm{eq}}\rightarrow\rho_S^h$. (B) Hot isochoric thermalization: $\rho_S^h\rightarrow\rho_S^{h,\mathrm{eq}}$. (C) Adiabatic expansion: $\rho_S^{h,\mathrm{eq}}\rightarrow\rho_S^c$. (D) Cold isochoric thermalization: $\rho_S^c\rightarrow\rho_S^{c,\mathrm{eq}}$.}
\end{figure}

\subsection{Process}

\textit{Process A: adiabatic compression}---In this process, the system does not interact with the reservoirs.
The Hamiltonian of the system is initialized to $H_S^c = \sum_iE_i|\psi_i\rangle\langle\psi_i|$, where each $E_i$ is an energy eigenvalue of $H_S^c$, and $|\psi_i\rangle$ is the corresponding eigenvector.
We assume that no degeneracy occurs in the eigenvalues and consider the initial state of the system to be the Gibbs state at inverse temperature $\beta_c$:
\begin{equation}
  \label{rho_s_c_eq}
  \rho_S^{c,\mathrm{eq}} = \frac{e^{-\beta_cH_S^c}}{Z_S^c} = \sum_i p_i|\psi_i\rangle\langle\psi_i|,
\end{equation}
where $Z_S^c = \mathrm{Tr}[e^{-\beta_cH_S^c}]$ and $p_i = e^{-\beta_cE_i}/Z_S^c$.

$H_S$ is dependent on a controllable external parameter $\lambda$. 
The initial value of $\lambda$ is $\lambda_i$, which corresponds to $H_S^c$.
During this process, $\lambda$ is varied from $\lambda_i$ to $\lambda_f$, and consequently, $H_S$ changes from $H_S^c$ to $H_S^h$.
The state change of the system in this process can be expressed by a unitary operator $U_\mathrm{com} = \mathcal{T}\exp\{ -i\int H_S(t)dt \}$, where $\mathcal{T}$ is the time-ordering operator.
If the change in $\lambda$ is sufficiently slow, the time-evolution induced by $U_\mathrm{com}$ does not change the probability distribution $\{p_i\}_i$ \cite{yi2017single}.
The final state of the system can be expressed as
\begin{equation}
  \label{rho_s_h}
  \rho_S^h = U_\mathrm{com}\; \rho_S^{c,\mathrm{eq}}\; U_\mathrm{com}^\dagger = \sum_i p_i|\phi_i\rangle\langle\phi_i|,
\end{equation}
where $|\phi_i\rangle = U_\mathrm{com}|\psi_i\rangle$  and $|\phi_i\rangle$ is the eigenvector of $H_S^h$ corresponding to the energy eigenvalue $\epsilon_i$ of $H_S^h$.
Here, the eigenvalue $E_i$ of $H_S^c$ and the eigenvalue $\epsilon_i$ of $H_S^h$ have a one-to-one correspondence, and no reversal of the magnitude relationship between eigenvalues and no degeneracy during this process are assumed.

Because the system does not interact with the reservoirs, no heat flows into the system during this process.
Therefore, we regard the change in internal energy of the system as the work performed on the system, which is given by
\begin{equation}
  W_\mathrm{com} = \mathrm{Tr}[H_S^h\rho_S^h] - \mathrm{Tr}[H_S^c\rho_S^{c,\mathrm{eq}}].
\end{equation}

\textit{Process B: hot isochoric thermalization}---In this process, the Hamiltonian of the system is constant at $H_S^h$.
A weak interaction exists between the system and the hot reservoir.
The reservoir is assumed to be sufficiently large such that its state does not change throughout this process.
After a sufficiently long time, the state of the system converges to the Gibbs state at inverse temperature $\beta_h$ \cite{ding2018measurement}.
The final state of the system can be expressed as
\begin{equation}
  \label{rho_s_h_eq}
  \rho_S^{h,\mathrm{eq}} = \frac{e^{-\beta_hH_S^h}}{Z_S^h} = \sum_i q_i|\phi_i\rangle\langle\phi_i|,
\end{equation}
where $Z_S^h = \mathrm{Tr}[e^{-\beta_hH_S^h}]$ and $q_i = e^{-\beta_h\epsilon_i} / Z_S^h$.
Each $\epsilon_i$ is an eigenvalue of $H_S^h$ and $|\phi_i\rangle$ is the corresponding eigenvector, which is equal to that used in Eq.~\eqref{rho_s_h}.

Here, we note that there is a crucial approximation in this model that the interaction between the system and the reservoir is ignored.
More precisely, the final state of the system should be the Gibbs state, considering the interaction Hamiltonian $H_{SB}^h$.
However, in this model, the contribution of $H_{SB}^h$ is neglected by assuming that the interaction is sufficiently weak; that is
\begin{equation}
  \frac{e^{-\beta_h(H_S^h + H_B^h + H_{SB}^h)}}{Z} \simeq \frac{e^{-\beta_hH_S^h}}{Z_S^h}\otimes\frac{e^{-\beta_hH_B^h}}{Z_B^h} = \rho_S^{h,\mathrm{eq}}\otimes\rho_B^h.
\end{equation}

Because the Hamiltonian is constant throughout this process, the work performed on the system is equal to 0.
Therefore, the change in internal energy of the system can be regarded as the heat transferred from the reservoir to the system, which is given by
\begin{equation}
  Q_\mathrm{in} = \mathrm{Tr}[H_S^h\rho_S^{h, \mathrm{eq}}] - \mathrm{Tr}[H_S^h\rho_S^h].
  \label{Qin_def}
\end{equation}

\textit{Process C: adiabatic expansion}---Similar to Process A, in Process C, no interaction occurs between the system and the reservoirs.
The parameter $\lambda$ is varied from $\lambda_f$ to $\lambda_i$ sufficiently slowly.
Consequently, the Hamiltonian of the system changes from $H_S^h$ to $H_S^c$.
$U_\mathrm{exp}$, the time-evolution operator of this process, is equal to $U_\mathrm{com}^\dagger$ \cite{anka2021measurement}.
Therefore, the final state can be expressed as follows:
\begin{equation}
  \rho_S^c = U_{\mathrm{exp}}\;\rho_S^{h, \mathrm{eq}}\;U_{\mathrm{exp}}^\dagger  = \sum_i q_i|\psi_i\rangle\langle\psi_i|,
\end{equation}
where $q_i$ is equal to that in Eq.~\eqref{rho_s_h_eq} and $|\psi_i\rangle$ is equal to that in Eq.~\eqref{rho_s_c_eq}.

Since the system does not interact with the reservoirs, no heat flows into the system and the work performed on the system is equal to the change in internal energy of the system; this is expressed as
\begin{equation}
  W_\mathrm{exp} = \mathrm{Tr}[H_S^c\rho_S^c] - \mathrm{Tr}[H_S^h\rho_S^{h, \mathrm{eq}}].
\end{equation}

\textit{Process D: cold isochoric thermalization}---In this process, the Hamiltonian of the system is constant at $H_S^c$ and the system weakly interacts with the cold reservoir.
Similar to Process B, after a sufficiently long time, the state of the system converges to the Gibbs state at inverse temperature $\beta_c$, i.e., the final state is $\rho_S^{c,\mathrm{eq}}$.
Here, the contribution of the interaction Hamiltonian $H_{SB}^c$ is neglected.

Because the Hamiltonian does not change during this process, the work performed on the system is 0 and the decrease in internal energy of the system is considered as the heat transferred into the cold reservoir, which is given by
\begin{equation}
  Q_{\mathrm{out}} = \mathrm{Tr}[H_S^c\rho_S^c] - \mathrm{Tr}[H_S^c\rho_S^{c, \mathrm{eq}}].
\end{equation}

\subsection{Efficiency}
With $Q_\mathrm{in}$ [Eq.~\eqref{Qin_def}] and $W_{\mathrm{out}}=-(W_{\mathrm{com}}+W_{\mathrm{exp}})$, the heat absorbed by the system from the hot reservoir and the net work performed by the system during one cycle, the efficiency of the weak coupling model is defined as
\begin{equation}
  \eta_{\mathrm{weak}} = \frac{W_{\mathrm{out}}}{Q_{\mathrm{in}}}.
\end{equation}
For the cycle to operate as a heat engine, we assume $Q_\mathrm{in} > Q_\mathrm{out} > 0$.
Because $W_\mathrm{com} + Q_\mathrm{in} + W_\mathrm{exp} - Q_\mathrm{out} = 0$, we can rewrite $\eta_\mathrm{weak}$ as
\begin{equation}
  \eta_{\mathrm{weak}} = \frac{Q_{\mathrm{in}} - Q_{\mathrm{out}}}{Q_{\mathrm{in}}} = 1 - \frac{Q_{\mathrm{out}}}{Q_{\mathrm{in}}}.
\end{equation}
Using the von Neumann entropy $S(\rho) \coloneqq -\mathrm{Tr}[\rho\ln\rho]$ and the quantum relative entropy $D(\rho||\sigma) \coloneqq \mathrm{Tr}[\rho\ln\rho] - \mathrm{Tr}[\rho\ln\sigma]$, the heat transferred between the system and the reservoirs can be expressed as follows (see Appendix A):
\begin{equation}
  \label{q_in}
  \beta_hQ_{\mathrm{in}} = \Delta S - D(\rho_S^h || \rho_S^{h, \mathrm{eq}}),
\end{equation}
\begin{equation}
  \label{q_out}
  \beta_cQ_{\mathrm{out}} = \Delta S + D(\rho_S^c || \rho_S^{c, \mathrm{eq}}),
\end{equation}
where $\Delta S = S(\rho_S^{h, \mathrm{eq}}) - S(\rho_S^h) = S(\rho_S^c) - S(\rho_S^{c, \mathrm{eq}})$.
Consequently, we can express $\eta_\mathrm{weak}$ as
\begin{equation}
  \label{eta_weak}
  \eta_{\mathrm{weak}} = 1 - \frac{\beta_h}{\beta_c}\frac{\Delta S + D(\rho_S^c || \rho_S^{c, \mathrm{eq}})}{\Delta S - D(\rho_S^h || \rho_S^{h, \mathrm{eq}})}.
\end{equation}
Because the terms expressed by the quantum relative entropy are positive under the condition $Q_\mathrm{in} > Q_\mathrm{out} > 0$, the following inequality holds:
\begin{equation}
  \eta_\mathrm{weak} < 1 - \frac{\beta_h}{\beta_c} = \eta_C.
\end{equation}
Therefore, the efficiency of the weak coupling model does not exceed $\eta_C$, which is the efficiency limit of classical heat engines.

\section{Strong coupling model}

In the weak coupling model, the interaction Hamiltonians are assumed to be negligible.
However, in quantum systems, this assumption is often unrealistic because the surface area of such systems is not insignificant compared with their volume \cite{strasberg2016nonequilibrium,perarnau2018strong}.
In this section, we develop a quantum Otto cycle model without applying the weak coupling assumption.
We call this the ``strong coupling model'' to distinguish it from the weak coupling model.

The strong coupling model consists of the following six processes: (A) adiabatic compression, (B-1) hot isochoric thermalization, (B-2) decoupling from the hot reservoir, (C) adiabatic expansion, (D-1) cold isochoric thermalization, and (D-2) decoupling from the cold reservoir.
Despite a few differences in details, this division is similar to the model proposed in Ref.~\cite{newman2017performance}.

\begin{figure}
\includegraphics[width=8.5cm]{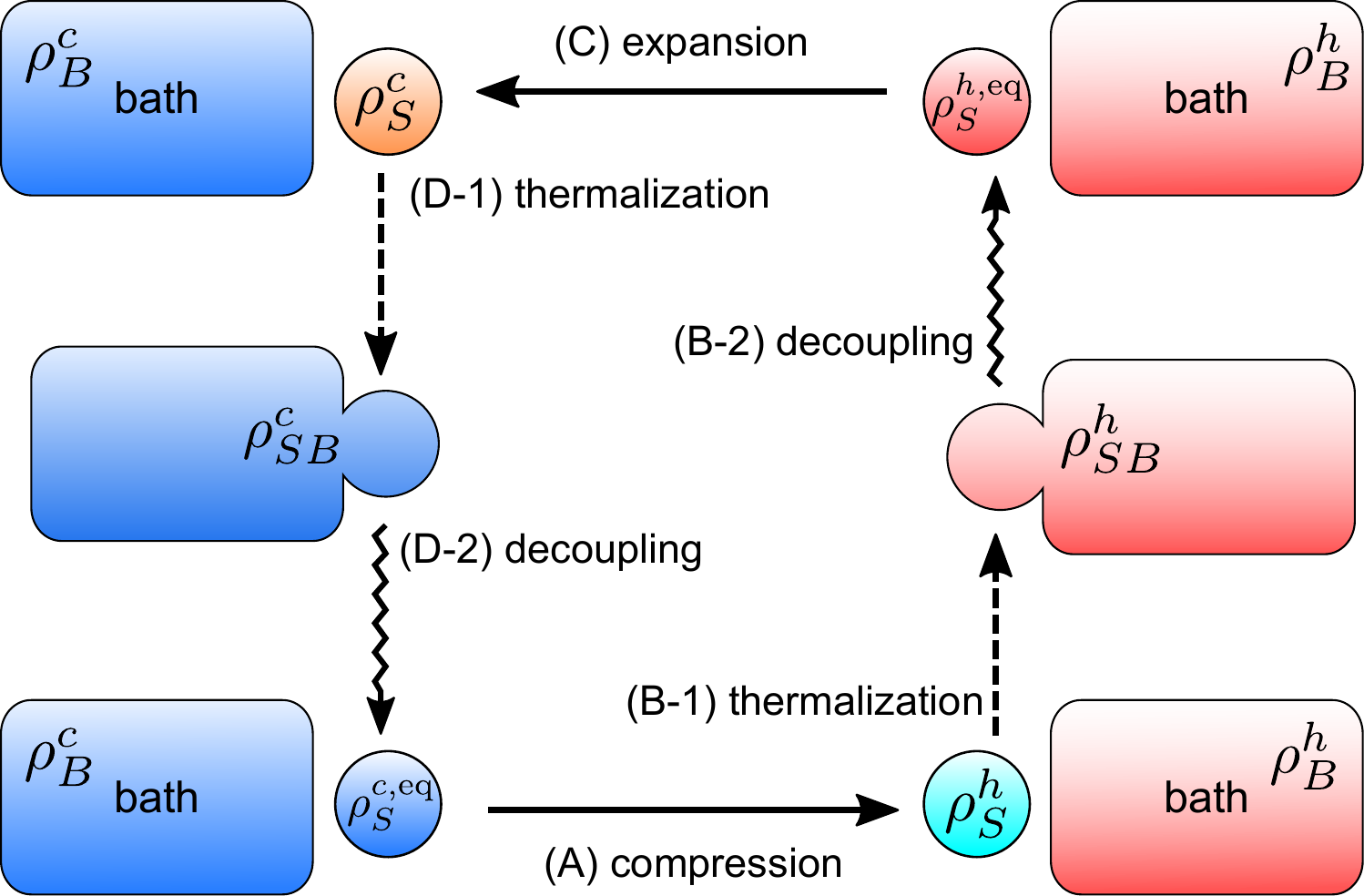}
  \caption{Six processes in the strong coupling model. While entangling with the reservoirs as appropriate, the state of the system transitions as follows. (A) Adiabatic compression: $\rho_S^{c,\mathrm{eq}}\rightarrow\rho_S^h$. (B-1) Hot isochoric thermalization: $\rho_S^h\otimes\rho_B^h\rightarrow\rho_{SB}^h$. (B-2) Decoupling from hot reservoir: $\rho_{SB}^h\rightarrow\rho_S^{h,\mathrm{eq}}\otimes\rho_B^h$. (C) Adiabatic expansion: $\rho_S^{h,\mathrm{eq}}\rightarrow\rho_S^c$. (D-1) Cold isochoric thermalization: $\rho_S^c\otimes\rho_B^c\rightarrow\rho_{SB}^c$. (D-2) Decoupling from cold reservoir: $\rho_{SB}^c\rightarrow\rho_S^{c,\mathrm{eq}}\otimes\rho_B^c$.}
\end{figure}

\subsection{Process}

\textit{Process A: adiabatic compression}---This process is equivalent to Process A in the weak coupling model.
No interaction occurs between the system and the reservoirs.
Therefore, no heat flows into the system.
The state of the system changes from $\rho_S^{c,\mathrm{eq}}$ to $\rho_S^h$, and the work $W_\mathrm{com} = \mathrm{Tr}[H_S^h\rho_S^h] - \mathrm{Tr}[H_S^c\rho_S^{c, \mathrm{eq}}]$ is done on the system during the process.

\textit{Process B-1: hot isochoric thermalization}---In this process, the Hamiltonian of the system is constant at $H_S^h$.
First, the system is coupled to the hot reservoir, which is represented by switching on the interaction Hamiltonian $H_{SB}^h$.
We assume that this operation does not require work because the reservoir is in Gibbs state initially.
We also assume the coupling operation completes instantaneously to ensure that the states of the system and the reservoir do not change before and after coupling.
These assumptions are also accepted in Refs. \cite{newman2017performance, newman2020quantum}.
From these assumptions, $\mathrm{Tr}[H_{SB}^h(\rho_S^h\otimes\rho_B^h)] = 0$ holds.
$H_{SB}^h$ is the Hamiltonian corresponding to the interaction between the system and the hot reservoir, and is constant in this process.
We note that $H_{SB}^h$ is not necessarily weak, which is a difference from the weak coupling model.
After a sufficiently long time, the compound system $S+B_h$ converges to the Gibbs state at inverse temperature $\beta_h$ \cite{rivas2020strong, newman2017performance}.
The final state can be expressed as follows:
\begin{equation}
  \rho_{SB}^h = \frac{e^{-\beta_h(H_S^h + H_B^h + H_{SB}^h)}}{Z_{SB}^h},
\end{equation}
where $Z_{SB}^h = \mathrm{Tr}[e^{-\beta_h(H_S^h + H_B^h + H_{SB}^h)}]$.

Similar to Process B of the weak coupling model, the work performed on the system vanishes.
Therefore, the change in internal energy of the system is equal to the heat transferred from the reservoir, which is
\begin{align}
  Q_\mathrm{th}^h &= \mathrm{Tr}[(H_S^h\otimes\mathbb{I}_B)\rho_{SB}^h] - \mathrm{Tr}[H_S^h\rho_S^h] + \mathrm{Tr}[H_{SB}^h\rho_{SB}^h] \notag \\
  &= \mathrm{Tr}[H_S^h(\tilde{\rho}_S^h - \rho_S^h)] + \mathrm{Tr}[H_{SB}^h\rho_{SB}^h].
\end{align}
Here, $\mathbb{I}_B$ is the identity operator and $\tilde{\rho}_S^h$ is the reduced state of $\rho_{SB}^h$, i.e., $\tilde{\rho}_S^h = \mathrm{Tr}_B[\rho_{SB}^h]$.
Using $Q_\mathrm{th}^h$, the work performed on the system can be written as
\begin{equation}
  W_\mathrm{th}^h = \mathrm{Tr}[H_S^h(\tilde{\rho}_S^h - \rho_S^h)] + \mathrm{Tr}[H_{SB}^h\rho_{SB}^h] - Q_\mathrm{th}^h = 0.
\end{equation}
With the von Neumann entropy and the quantum relative entropy, we can rewrite $Q_\mathrm{th}^h$ as follows (see Appendix A):
\begin{align}
    \label{q_th_h}
    \beta_{h}Q_{\mathrm{th}}^{h}=&S(\tilde{\rho}_{S}^{h})-S(\rho_{S}^{h})\notag\\&-\left\{ D(\rho_{SB}^{h}||\tilde{\rho}_{S}^{h}\otimes\rho_{B}^{h})+D(\rho_{S}^{h}\otimes\rho_{B}^{h}||\rho_{SB}^{h})\right\} .
\end{align}
This expression is useful for the calculation of the efficiency.

\textit{Process B-2: decoupling from hot reservoir}---Next, we conduct an operation to detach the system from the hot reservoir.
As in Ref.~\cite{xu2018achieving}, we consider the situation where the decoupling and the thermalization proceed simultaneously, i.e., the decoupling operation is sufficiently slow and the state of the compound system $S+B_h$ is the Gibbs state throughout this process.
The initial state is $\rho_{SB}^h$ and the final state is $\rho_S^{h,\mathrm{eq}}\otimes\rho_B^h$.
Here, we impose a restriction that this process can be realized under the Schr\"odinger equation.
From this restriction, the time evolution of the compound system is unitary and the following equality holds:
\begin{equation}
  S(\rho_S^{h,\mathrm{eq}}\otimes\rho_B^h) = S(\rho_{SB}^h).
\end{equation}
Hereafter, we refer to this restriction as ``unitary restriction.''
We point out that the interaction Hamiltonian is not constant during this process; its initial value is $H_{SB}^h$ and the final value is 0, and it varies appropriately to realize the desired unitary transformation.

The appropriate definition of work and heat is an open question in the field of quantum thermodynamics and the efficiency of quantum heat engine is largely dependent the definition.
In this study, we introduce the definition of heat proposed in Ref.~\cite{xu2018achieving}, which is valid when the compound system evolves under the Schr\"{o}dinger equation.
Using this definition, the heat transferred into the system can be calculated as follows (see Appendix B):
\begin{equation}
  \label{q_d_h}
  \beta_hQ_d^h = S(\rho_S^{h,\mathrm{eq}}) - S(\tilde{\rho}_S^h) + D(\rho_{SB}^h||\tilde{\rho}_S^h\otimes\tilde{\rho}_B^h),
\end{equation}
where $\tilde{\rho}_B^h = \mathrm{Tr}_S[\rho_{SB}^h]$.
We define the work performed on the system as the difference between the change in internal energy of the system and $Q_d^h$, which is
\begin{equation}
  W_d^h = \mathrm{Tr}[H_S^h(\rho_S^{h,\mathrm{eq}} - \tilde{\rho}_S^h)] - \mathrm{Tr}[H_{SB}^h\rho_{SB}^h] - Q_d^h.
\end{equation}

We note that the state change of the system $\rho_S^h \rightarrow \rho_S^{h,\mathrm{eq}}$ involves both work and heat in the strong coupling model, whereas only heat transfer occurs in the weak coupling model.
This is a fundamental difference between both model.

\textit{Process C: adiabatic expansion}---This process is completely equivalent to Process C of the weak coupling model.
The state of the system changes from $\rho_S^{h,\mathrm{eq}}$ to $\rho_S^c$ without interaction with the reservoirs. No heat flows into the system and the work performed on the system is $W_\mathrm{exp} = \mathrm{Tr}[H_S^c\rho_S^c] - \mathrm{Tr}[H_S^h\rho_S^{h,\mathrm{eq}}]$.

\textit{Process D-1: cold isochoric thermalization}---In this process, the Hamiltonian of the system is constant at $H_S^c$.
First, the system is coupled to the cold reservoir, and we assume $\mathrm{Tr}[H_{SB}^c(\rho_S^c\otimes\rho_B^c)]=0$ as in Process B-1.
$H_{SB}^c$ corresponds to the interaction between the system and the cold reservoir, and it is constant in this process.
We emphasize that $H_{SB}^c$ is not necessarily negligible.
After a sufficiently long time, the state of the compound system $S+B_c$ transitions to the Gibbs state at inverse temperature $\beta_c$.
The final state is
\begin{equation}
  \rho_{SB}^c = \frac{e^{-\beta_c(H_S^c + H_B^c + H_{SB}^c)}}{Z_{SB}^c},
\end{equation}
where $Z_{SB}^c = \mathrm{Tr}[e^{-\beta_c(H_S^c + H_B^c + H_{SB}^c)}]$.

Similar to Process B-1, the work performed on the system is 0.
Therefore, the change in internal energy of the system can be regarded as $Q_\mathrm{th}^c$, the heat transferred into the system during this process.
$Q_\mathrm{th}^c$ is given by
\begin{align}
  Q_\mathrm{th}^c &= \mathrm{Tr}[(H_S^c\otimes\mathbb{I}_B)\rho_{SB}^c] - \mathrm{Tr}[H_S^c\rho_S^c] + \mathrm{Tr}[H_{SB}^c\rho_{SB}^c] \notag \\
  &= \mathrm{Tr}[H_S^c(\tilde{\rho}_S^c - \rho_S^c)] + \mathrm{Tr}[H_{SB}^c\rho_{SB}^c],
\end{align}
where $\tilde{\rho}_S^c = \mathrm{Tr}_B[\rho_{SB}^c]$.
We note that $Q_\mathrm{th}^c$ is calculated with the flow into the system in a positive direction, despite the positive energy actually being transferred from the system to the cold reservoir.
Using $Q_\mathrm{th}^c$, the work performed on the system can be written as
\begin{equation}
  W_\mathrm{th}^c = \mathrm{Tr}[H_S^c(\tilde{\rho}_S^c - \rho_S^c)] + \mathrm{Tr}[H_{SB}^c\rho_{SB}^c] - Q_\mathrm{th}^c = 0.
\end{equation}
Similar to $Q_\mathrm{th}^h$, another expression for $Q_\mathrm{th}^c$ can be obtained as follows:
\begin{align}
    \label{q_th_c}
    \beta_{c}Q_{\mathrm{th}}^{c}=&S(\tilde{\rho}_{S}^{c})-S(\rho_{S}^{c})\notag\\&-\left\{ D(\rho_{SB}^{c}||\tilde{\rho}_{S}^{c}\otimes\rho_{B}^{c})+D(\rho_{S}^{c}\otimes\rho_{B}^{c}||\rho_{SB}^{c})\right\} .
\end{align}

\textit{Process D-2: decoupling from cold reservoir}---Similar to Process B-2, in this process, the system is decoupled from the cold reservoir sufficiently slowly, and the thermalization proceeds simultaneously.
The compound system $S+B_c$ is the Gibbs state at inverse temperature $\beta_c$ throughout this process.
The final state is $\rho_S^{c,\mathrm{eq}}\otimes\rho_B^c$.
Here, we impose the unitary restriction as in Process B-2, i.e., this process can be realized under the Schr\"odinger equation and the following equality holds:
\begin{equation}
  S(\rho_S^{c,\mathrm{eq}}\otimes\rho_B^c) = S(\rho_{SB}^c).
\end{equation}

By adopting the same definition of heat as in Process B-2, the heat the system absorbs from the cold reservoir can be calculated as follows:
\begin{equation}
  \label{q_d_c}
  \beta_cQ_d^c = S(\rho_S^{c,\mathrm{eq}}) - S(\tilde{\rho}_S^c) + D(\rho_{SB}^c||\tilde{\rho}_S^c\otimes\tilde{\rho}_B^c),
\end{equation}
where $\tilde{\rho}_B^c = \mathrm{Tr}_S[\rho_{SB}^c]$.
The work performed on the system is defined as the difference between the change in internal energy of the system and $Q_d^c$:
\begin{equation}
  W_d^c = \mathrm{Tr}[H_S^c(\rho_S^{c,\mathrm{eq}} - \tilde{\rho}_S^c)] - \mathrm{Tr}[H_{SB}^c\rho_{SB}^c] - Q_d^c.
\end{equation}

\subsection{Efficiency}

Similar to the weak coupling model, we define the efficiency of the strong coupling model as
\begin{equation}
  \eta_\mathrm{str} = \frac{W_\mathrm{out}'}{Q_\mathrm{in}'},
\end{equation}
where $Q_\mathrm{in}'$ is the sum of the heat the system absorbs from the hot reservoir in Process B-1 and Process B-2 and
$W_\mathrm{out}'$ is the net work the system performs during one cycle.
Hence, $W_\mathrm{out}' = -(W_\mathrm{com} + W_\mathrm{th}^h + W_d^h + W_\mathrm{exp} + W_\mathrm{th}^c + W_d^c)$ and $Q_\mathrm{in}' = Q_\mathrm{th}^h + Q_d^h$.
Additionally, $Q_\mathrm{out}' = -(Q_\mathrm{th}^c + Q_d^c)$ denotes the heat transferred from the system to the cold reservoir.
Similar to the weak coupling model, we assume $Q_\mathrm{in}' > Q_\mathrm{out}' > 0$.
We can derive
\begin{align}
  W_\mathrm{out}' &= Q_\mathrm{th}^h + Q_d^h + Q_\mathrm{th}^c + Q_d^c \notag \\
  &= Q_\mathrm{in}' - Q_\mathrm{out}'
\end{align}
and we can rewrite $\eta_\mathrm{str}$ as
\begin{equation}
  \eta_\mathrm{str} = \frac{Q_\mathrm{in}' - Q_\mathrm{out}'}{Q_\mathrm{in}'} = 1 - \frac{Q_\mathrm{out}'}{Q_\mathrm{in}'}.
\end{equation}
Moreover, from Eqs.~\eqref{q_th_h} and \eqref{q_d_h}, $Q_\mathrm{in}'$ can be calculated as follows:
\begin{widetext}
\begin{align}
  \beta_hQ_\mathrm{in}' &= S(\rho_S^{h,\mathrm{eq}}) - S(\rho_S^h) - \left\{ D(\rho_{SB}^h||\tilde{\rho}_S^h\otimes\rho_B^h) - D(\rho_{SB}^h||\tilde{\rho}_S^h\otimes\tilde{\rho}_B^h) + D(\rho_S^h\otimes\rho_B^h||\rho_{SB}^h) \right\} \notag \\
   &= S(\rho_S^{h,\mathrm{eq}}) - S(\rho_S^h)  - \left\{ D(\tilde{\rho}_B^h||\rho_B^h) + D(\rho_S^h\otimes\rho_B^h||\rho_{SB}^h) \right\}.
\end{align}
Here, we used the following relation:
\begin{align}
  D(\rho_{SB}^h||\tilde{\rho}_S^h\otimes\rho_B^h) - D(\rho_{SB}^h||\tilde{\rho}_S^h\otimes\tilde{\rho}_B^h) &= \left\{ -S(\rho_{SB}^h) +S(\tilde{\rho}_S^h) -\mathrm{Tr}[\tilde{\rho}_B^h\ln\rho_B^h] \right\} - \left\{ -S(\rho_{SB}^h) +S(\tilde{\rho}_S^h) -\mathrm{Tr}[\tilde{\rho}_B^h\ln\tilde{\rho}_B^h] \right\} \notag \\
  &= \mathrm{Tr}[\tilde{\rho}_B^h\ln\tilde{\rho}_B^h] - \mathrm{Tr}[\tilde{\rho}_B^h\ln\rho_B^h] \notag \\
  &= D(\tilde{\rho}_B^h||\rho_B^h).
\end{align}
\end{widetext}
Similarly, from Eqs.~\eqref{q_th_c} and \eqref{q_d_c}, we obtain
\begin{align}
  \beta_cQ_\mathrm{out}' = &S(\rho_S^c) - S(\rho_S^{c,\mathrm{eq}}) \notag \\  &+ \left\{ D(\tilde{\rho}_B^c||\rho_B^c) + D(\rho_S^c\otimes\rho_B^c||\rho_{SB}^c) \right\}.
\end{align}
Thus, using $\Delta S = S(\rho_S^{h,\mathrm{eq}}) -  S(\rho_S^h) = S(\rho_S^c) - S(\rho_S^{c,\mathrm{eq}})$, $\eta_\mathrm{str}$ can be expressed as
\begin{equation}
  \label{eta_str}
  \eta_\mathrm{str} = 1 - \frac{\beta_h}{\beta_c}\frac{\Delta S + \left\{ D(\tilde{\rho}_B^c||\rho_B^c) + D(\rho_S^c\otimes\rho_B^c||\rho_{SB}^c) \right\} }{\Delta S - \left\{ D(\tilde{\rho}_B^h||\rho_B^h) + D(\rho_S^h\otimes\rho_B^h||\rho_{SB}^h) \right\} }.
\end{equation}
This expression resembles $\eta_\mathrm{weak}$ in Eq.~\eqref{eta_weak}.
Equation ~\eqref{eta_str} is the main result of this study.
From the non-negativity of the quantum relative entropy and the condition $Q_\mathrm{in}' > Q_\mathrm{out}' > 0$, we obtain
\begin{equation}
  \eta_\mathrm{str} < 1 - \frac{\beta_h}{\beta_c} = \eta_C,
\end{equation}
i.e., similar to the weak coupling model, the efficiency of our strong coupling model does not exceed the classical limit $\eta_C$.

\subsection{Weak coupling limit}
Here, we show that our strong coupling model agrees with the weak coupling model in the limit of weak interaction.
When we add a condition that the interaction Hamiltonians are negligible in the strong coupling model, the final state of Process B-1 is approximated as follows:
\begin{align}
  \rho_{SB}^h &= \frac{e^{-\beta_h(H_S^h + H_B^h + H_{SB}^h)}}{Z_{SB}^h} \notag \\
  & \simeq \frac{e^{-\beta_h(H_S^h + H_B^h)}}{Z_S^hZ_B^h} = \frac{e^{-\beta_hH_S^h}}{Z_S^h}\otimes\frac{e^{-\beta_hH_B^h}}{Z_B^h} = \rho_S^{h,\mathrm{eq}}\otimes\rho_B^h.
\end{align}
Consequently, $\tilde{\rho}_S^h = \mathrm{Tr}_B[\rho_{SB}^h] = \rho_S^{h,\mathrm{eq}}$ and $\tilde{\rho}_B^h = \mathrm{Tr}_S[\rho_{SB}^h] = \rho_B^h$ hold.
Therefore, we can rewrite the heat exchanged between the system and the hot reservoir as
\begin{equation}
  \label{q_th_h_limit}
  Q_\mathrm{th}^h = \mathrm{Tr}[H_S^h(\tilde{\rho}_S^h-\rho_S^h)] + \mathrm{Tr}[H_{SB}^h\rho_{SB}^h] \simeq \mathrm{Tr}[H_S^h(\rho_S^{h,\mathrm{eq}}-\rho_S^h)],
\end{equation}
\begin{align}
  \label{q_d_h_limit}
  \beta_hQ_d^h &= S(\rho_S^{h,\mathrm{eq}}) - S(\tilde{\rho}_S^h) + D(\rho_{SB}^h||\tilde{\rho}_S^h\otimes\tilde{\rho}_B^h) \notag \\
  &\simeq S(\rho_S^{h,\mathrm{eq}}) - S(\rho_S^{h,\mathrm{eq}}) + D(\rho_S^{h,\mathrm{eq}}\otimes\rho_B^h||\rho_S^{h,\mathrm{eq}}\otimes\rho_B^h) \notag \\
  &= 0.
\end{align}
From Eqs.~\eqref{q_th_h_limit} and \eqref{q_d_h_limit}, we obtain $Q_\mathrm{in}' = Q_\mathrm{th}^h + Q_d^h \simeq \mathrm{Tr}[H_S^h(\rho_S^{h,\mathrm{eq}}-\rho_S^h)] = Q_\mathrm{in}$.
By similar calculations, we can show $Q_\mathrm{out}' \simeq Q_\mathrm{out}$ and $W_\mathrm{out}' \simeq W_\mathrm{out}$.
These relations mean the cycle of the strong coupling model agrees with that of the weak coupling model in the limit of weak interaction.
Furthermore, the following calculation shows that $\eta_\mathrm{str}$ in Eq.~\eqref{eta_str} agrees with $\eta_\mathrm{weak}$ in the same limit:
\begin{align}
  \eta_\mathrm{str} &= 1 - \frac{\beta_h}{\beta_c}\frac{\Delta S + \left\{ D(\tilde{\rho}_B^c||\rho_B^c) + D(\rho_S^c\otimes\rho_B^c||\rho_{SB}^c)  \right\} }{\Delta S - \left\{ D(\tilde{\rho}_B^h||\rho_B^h) + D(\rho_S^h\otimes\rho_B^h||\rho_{SB}^h) \right\} } \notag \\
  &\simeq 1 - \frac{\beta_h}{\beta_c}\frac{\Delta S + \left\{ D(\rho_B^c||\rho_B^c) + D(\rho_S^c\otimes\rho_B^c||\rho_S^{c,\mathrm{eq}}\otimes\rho_B^c) \right\} }{\Delta S - \left\{ D(\rho_B^h||\rho_B^h) + D(\rho_S^h\otimes\rho_B^h||\rho_S^{h,\mathrm{eq}}\otimes\rho_B^h) \right\} } \notag \\
  &= 1 - \frac{\beta_h}{\beta_c}\frac{\Delta S + D(\rho_S^c||\rho_S^{c,\mathrm{eq}})}{\Delta S - D(\rho_S^h||\rho_S^{h,\mathrm{eq}})} \notag \\
  &= \eta_\mathrm{weak}.
\end{align}
These discussions support the consistency of the proposed strong coupling model with the weak coupling model.
Thus, our model can be considered to be a valid extension of the existing weak coupling model.

\subsection{Comparison of efficiencies}
In this subsection, we compare the efficiency of the two models discussed previously.
From Eqs.~\eqref{eta_weak} and \eqref{eta_str}, we derive the sufficient condition for $\eta_\mathrm{str} \leq \eta_\mathrm{weak}$.
First, from the non-negativity of the quantum relative entropy, we obtain $D(\tilde{\rho}_B^h||\rho_B^h) + D(\rho_S^h\otimes\rho_B^h||\rho_{SB}^h) \geq D(\rho_S^h\otimes\rho_B^h||\rho_{SB}^h)$.
We set $d\coloneqq D(\rho_S^h\otimes\rho_B^h||\rho_{SB}^h) - D(\rho_S^h||\rho_S^{h,\mathrm{eq}})$, and by examining the sign of $d$ as the following calculation, we explore the magnitude relation between the efficiencies of the two models:
\begin{widetext}
\begin{align}
  d &\coloneqq D(\rho_S^h\otimes\rho_B^h||\rho_{SB}^h) - D(\rho_S^h||\rho_S^{h,\mathrm{eq}}) \notag \\
  &= -S(\rho_S^h\otimes\rho_B^h) -\mathrm{Tr}\left[ (\rho_S^h\otimes\rho_B^h)\ln\frac{e^{-\beta_h(H_S^h+H_B^h+H_{SB}^h)}}{Z_{SB}^h} \right] + S(\rho_S^h) + \mathrm{Tr}\left[ \rho_S^h\ln\frac{e^{-\beta_hH_S^h}}{Z_S^h} \right] \notag \\
  &= \ln\frac{Z_{SB}^h}{Z_S^hZ_B^h} \notag \\
  &= \beta_h\left\{ F(\rho_S^{h,\mathrm{eq}}\otimes\rho_B^h) - F(\rho_{SB}^h) \right\}.
\end{align}
\end{widetext}
Here, we introduced the free energy $F(\rho) \coloneqq -\frac{1}{\beta}\ln Z$, defined for an arbitrary Gibbs state.
$\beta$ is the inverse temperature and $Z$ is the partition function.
Thus, if $F(\rho_S^{h,\mathrm{eq}}\otimes\rho_B^h) \geq F(\rho_{SB}^h)$ holds, we obtain $d\geq 0$ and $D(\tilde{\rho}_B^h||\rho_B^h) + D(\rho_S^h\otimes\rho_B^h||\rho_{SB}^h) \geq D(\rho_S^h||\rho_S^{h,\mathrm{eq}})$.
Similarly, we also obtain $D(\tilde{\rho}_B^c||\rho_B^c) + D(\rho_S^c\otimes\rho_B^c||\rho_{SB}^c) \geq D(\rho_S^c||\rho_S^{c,\mathrm{eq}})$ under the condition that $F(\rho_S^{c,\mathrm{eq}}\otimes\rho_B^c) \geq F(\rho_{SB}^c)$.
From the two inequalities, we can derive the following relation:
\begin{align}
  \eta_\mathrm{str} &= 1 - \frac{\beta_h}{\beta_c}\frac{\Delta S + \left\{ D(\tilde{\rho}_B^c||\rho_B^c) + D(\rho_S^c\otimes\rho_B^c||\rho_{SB}^c) \right\} }{\Delta S - \left\{ D(\tilde{\rho}_B^h||\rho_B^h) + D(\rho_S^h\otimes\rho_B^h||\rho_{SB}^h) \right\} } \notag \\
  &\leq 1 - \frac{\beta_h}{\beta_c}\frac{\Delta S + D(\rho_S^c||\rho_S^{c,\mathrm{eq}})}{\Delta S - D(\rho_S^h||\rho_S^{h,\mathrm{eq}})} \notag \\
  &= \eta_\mathrm{weak}.
\end{align}
Eventually, the fact that both $F(\rho_S^{h,\mathrm{eq}}\otimes\rho_B^h) \geq F(\rho_{SB}^h)$ and $F(\rho_S^{c,\mathrm{eq}}\otimes\rho_B^c) \geq F(\rho_{SB}^c)$ are satisfied is the sufficient condition for $\eta_\mathrm{str} \leq \eta_\mathrm{weak}$.
Furthermore, because the von Neumann entropy of the compound system is constant throughout the decoupling processes due to the unitary restriction, the change in free energy is equal to the change in internal energy, i.e.,
\begin{align}
  \label{delta_f_h}
  &F(\rho_S^{h,\mathrm{eq}}\otimes\rho_B^h) - F(\rho_{SB}^h) \\
  &= \mathrm{Tr}[H_S^h(\rho_S^{h,\mathrm{eq}}-\tilde{\rho}_S^h)] + \mathrm{Tr}[H_B^h(\rho_B^h-\tilde{\rho}_B^h)] - \mathrm{Tr}[H_{SB}^h\rho_{SB}^h]\notag,
\end{align}
\begin{align}
  \label{delta_f_c}
  &F(\rho_S^{c,\mathrm{eq}}\otimes\rho_B^c) - F(\rho_{SB}^c) \\
  &= \mathrm{Tr}[H_S^c(\rho_S^{c,\mathrm{eq}}-\tilde{\rho}_S^c)] + \mathrm{Tr}[H_B^c(\rho_B^c-\tilde{\rho}_B^c)] - \mathrm{Tr}[H_{SB}^c\rho_{SB}^c]\notag.
\end{align}
Therefore, using $\Delta E_d^h$ and $\Delta E_d^c$ to denote the right sides of Eqs. \eqref{delta_f_h} and \eqref{delta_f_c}, the fact that both $\Delta E_d^h \geq 0$ and $\Delta E_d^c \geq 0$ are satisfied also expresses the sufficient condition for $\eta_\mathrm{str} \leq \eta_\mathrm{weak}$.
We can interpret this condition that if positive costs exist on the two decoupling processes, the strong coupling model has a lower efficiency than the weak coupling model.

\subsection{Reversal of efficiency}
This subsection shows that the efficiency of the strong coupling model may exceed that of the weak coupling model in particular case.
In this subsection, only high temperature reservoir side is discussed.
However, the same is true for the low temperature side.
An arbitrary interaction Hamiltonian can be expressed as follows:
\begin{equation}
    \label{Hsb}
    H_{SB}^h = U_d^{h\dagger} (H_S^h+H_B^h)U_d^h + a\mathbb{I} - (H_S^h+H_B^h),
\end{equation}
where the unitary matrix $U_d^h$ represents Process B-2 in the strong coupling model and $a = \mathrm{Tr}[(H_S^h+H_B^h)(\rho_S^h\otimes\rho_B^h)] - \mathrm{Tr}[(H_S^h+H_B^h)U_d^h(\rho_S^h\otimes\rho_B^h)U_d^{h\dagger}] $.
This interaction Hamiltonian satisfies the restriction $\mathrm{Tr}[H_{SB}^h(\rho_S^h\otimes\rho_B^h)]=0$ and the unitary restriction: the eigenvalue distribution of $H_S^h+H_B^h+H_{SB}^h$ is equal to that of $H_S^h+H_B^h$ except for the shift of the constant $a$.
The essential difference between the two models is whether the interaction Hamiltonians are considered.
Taking the interaction Hamiltonian into account is equivalent to considering any $U_d^h$ and transforming it by Eq. \eqref{Hsb}.

Next, we present the construction of $U_d^h$ which achieves the relation $d = D(\rho_S^h\otimes\rho_B^h||\rho_{SB}^h) - D(\rho_S^h||\rho_S^{h,\mathrm{eq}}) < 0$.
As described in the previous subsection, examining the sign of $d$ helps determine the magnitude relation between $\eta_\mathrm{weak}$ and $\eta_\mathrm{str}$, and the relation $d<0$ indicates that the efficiency of the strong coupling model can be higher than the weak coupling counterpart (compare Eq. \eqref{eta_str} with Eq. \eqref{eta_weak}).
We use the following notations: $\rho_B^h = \sum_iq_i'\ket{\phi_i'}\bra{\phi_i'}$, $\ket{\Phi_{ij}} = \ket{\phi_i}\otimes\ket{\phi_j'}$, and let $\{\epsilon_i\}_i$ and $\{\epsilon_i'\}_i$ to be the eigenvalues of $H_S^h$ and $H_B^h$, respectively.
The value of $d$ can be calculated as follows:
\begin{widetext}
\begin{align}
\label{d_value}
  d&\coloneqq D(\rho_S^h\otimes\rho_B^h||\rho_{SB}^h) - D(\rho_S^h||\rho_S^{h,\mathrm{eq}}) \notag \\
  &= D\left( U_d^h(\rho_S^h\otimes\rho_B^h)U_d^{h\dagger} || \rho_S^{h,\mathrm{eq}}\otimes\rho_B^h \right) - D(\rho_S^h\otimes\rho_B^h||\rho_S^{h,\mathrm{eq}}\otimes\rho_B^h) \notag \\
  &= -\mathrm{Tr}\left[ U_d^h(\rho_S^h\otimes\rho_B^h)U_d^{h\dagger}\ln(\rho_S^{h,\mathrm{eq}}\otimes\rho_B^h) \right] + \mathrm{Tr}\left[ (\rho_S^h\otimes\rho_B^h)\ln(\rho_S^{h,\mathrm{eq}}\otimes\rho_B^h) \right] \notag \\
&= \beta_h\sum_{i,j}\sum_{k,l}p_iq_j'(\epsilon_k+\epsilon_l') \left|\braket{\Phi_{kl}|U_d^h|\Phi_{ij}}\right|^2 - \beta_h\sum_{i,j}p_iq_j'(\epsilon_i+\epsilon_j').
\end{align}
\end{widetext}
For any real vectors $\boldsymbol{a}$ and $\boldsymbol{b}$, the following relation holds: $\boldsymbol{a^\top b} \geq \boldsymbol{a_\uparrow^\top b_\downarrow}$, where $\boldsymbol{a_\uparrow}$ and $\boldsymbol{b_\downarrow}$ are vectors of the elements of $\boldsymbol{a}$ sorted in ascending order and $\boldsymbol{b}$ sorted in descending order, respectively.
Here, let $\boldsymbol{E}$ be a vector of $(\epsilon_i+\epsilon_j')$ values ordered from smallest to largest and $\boldsymbol{P}$ be a vector of $p_iq_j'$ values, which is ordered corresponding to $\boldsymbol{E}$.
Using these vectors, the second term of Eq. \eqref{d_value} can be expressed as the inner product of them and the following holds:
\begin{equation}
    \sum_{i,j}p_iq_j'(\epsilon_i+\epsilon_j') = \boldsymbol{E^\top P} \geq \boldsymbol{E^\top P_\downarrow},
\end{equation}
where $\boldsymbol{P_\downarrow}$ is a vector of $\boldsymbol{P}$ in descending order.
Thus, if we take $U_d^h$ as the following form, $d<0$ is satisfied:
\begin{equation}
\label{Udh}
    U_d^h = \sum_{(m,n)\in\mathcal{M}}\ket{\Phi_n}\bra{\Phi_m},
\end{equation}
where $\mathcal{M}$ is the set of $(m,n)$, which is a correspondence of indices before and after $\boldsymbol{P}$ is sorted to $\boldsymbol{P}_\downarrow$.
This $U_d^h$ can be interpreted as a permutation matrix under the orthonormal basis $\{\ket{\Phi_n}\}_n$ and hence it is unitary.
Under this $U_d^h$, the value of $d$ is as follows:
\begin{equation}
    d = \beta_h \left( \boldsymbol{E^\top P_\downarrow} - \boldsymbol{E^\top P} \right) \leq 0.
\end{equation}
This mean that such $U_d^h$ can induce the reversal of efficiency, $\eta_\mathrm{str} > \eta_\mathrm{weak}$.

Furthermore, the $U_d^h$ constructed by Eq. \eqref{Udh} provides the minimum value of $d$.
For any $U_d^h$, the first term in Eq. \eqref{d_value} can be expressed as
\begin{equation}
    \beta_h\sum_{i,j}\sum_{k,l}p_iq_j'(\epsilon_k+\epsilon_l') \left|\braket{\Phi_{kl}|U_d^h|\Phi_{ij}}\right|^2 = \beta_h\boldsymbol{E^\top AP}.
\end{equation}
Here, $\boldsymbol{A} = (|\bra{\Phi_m}U_d^h\ket{\Phi_n}|^2)$ is a doubly stochastic matrix.
In general, a doubly stochastic matrix $\boldsymbol{A}$ generates majorization relation \cite{marshall1979inequalities,sagawa2022entropy}: for any real vector $x$, $\boldsymbol{A}x \prec x$.
From this property, we can rigorously prove the inequality
\begin{equation}
\label{majo_ineq}
    \boldsymbol{E^\top AP} \geq \boldsymbol{E^\top P_\downarrow}
\end{equation}
for any doubly stochastic matrix $\boldsymbol{A}$ (see Appendix C).
Therefore, the $U_d^h$ in Eq. \eqref{Udh} is optimal for minimizing the value of $d$.

This implies that the efficiency of the strong coupling model can exceed that of the weak coupling model, although this does not necessarily realize because of ignoring the contribution from the term $D(\tilde{\rho}_B^h||\rho_B^h)$.
In the following section, we demonstrate numerically that the reversal of the efficiency is achieved in the prepared simple two-level system.

\section{Example}
In this section, the strong coupling model is applied to a simple two-level system, and its efficiency is numerically computed.
We represent the state of the reservoirs by the general density operators and do not specify the details of the reservoirs.
We introduce a parameter $\theta$ to control the strength of the interaction and examine the relationship between $\theta$ and $\eta_\mathrm{str}$. 
In addition, we analytically calculate the efficiency of the weak coupling model and compare it with $\eta_\mathrm{str}$.
Furthermore, we design the interaction Hamiltonians by using Eqs. \eqref{Hsb} and \eqref{Udh} and investigate the efficiency of the strong coupling model under such interaction Hamiltonians.

Consider a two-level system $S$.
Using the computational basis $|0\rangle = (1,0)^\top,\;|1\rangle = (0,1)^\top$ and the other orthonormal basis $|\pm\rangle = \frac{1}{\sqrt{2}}(1, \pm 1)^\top$, we set $H_S^c$ and $H_S^h$ as
\begin{equation}
  H_S^c = E_g^c|0\rangle\langle 0| + E_e^c|1\rangle\langle 1|, \;\;H_S^h = E_g^h|+\rangle\langle +| + E_e^h|-\rangle\langle -|.
\end{equation}
$\{E_g^c, E_e^c\}$ and $\{E_g^h, E_e^h\}$ are the energy eigenvalues of $H_S^c$ and $H_S^h$, respectively.
Let $\{\epsilon_i^c\}_i$ and $\{\epsilon_i^h\}_i$ be the eigenvalues of the Hamiltonians of the cold and hot reservoirs, and these Hamiltonians are assumed to be diagonalized in the computational basis.
Then, we generate two unitary operators $U_j$ ($j=c,h$; same hereafter) randomly based on Haar measurement \cite{maris2009generate,mezzadri2006generate} and calculate the Hermitian operators $H_j$ such that $U_j = \exp(iH_j)$.
By introducing a parameter $\theta \in [0,1]$, we construct the interaction Hamiltonians by substituting $U_d^j = \exp(iH_j\theta)$ into Eq. \eqref{Hsb}.
These specify the strong coupling model and various quantities, such as heat and efficiency, can be calculated using them.
We note that the strong coupling model is equivalent to the weak coupling model when $\theta = 0$ and as $\theta$ increases, we can consider the model with stronger interactions.

\begin{figure}
  \includegraphics[width=8.5cm]{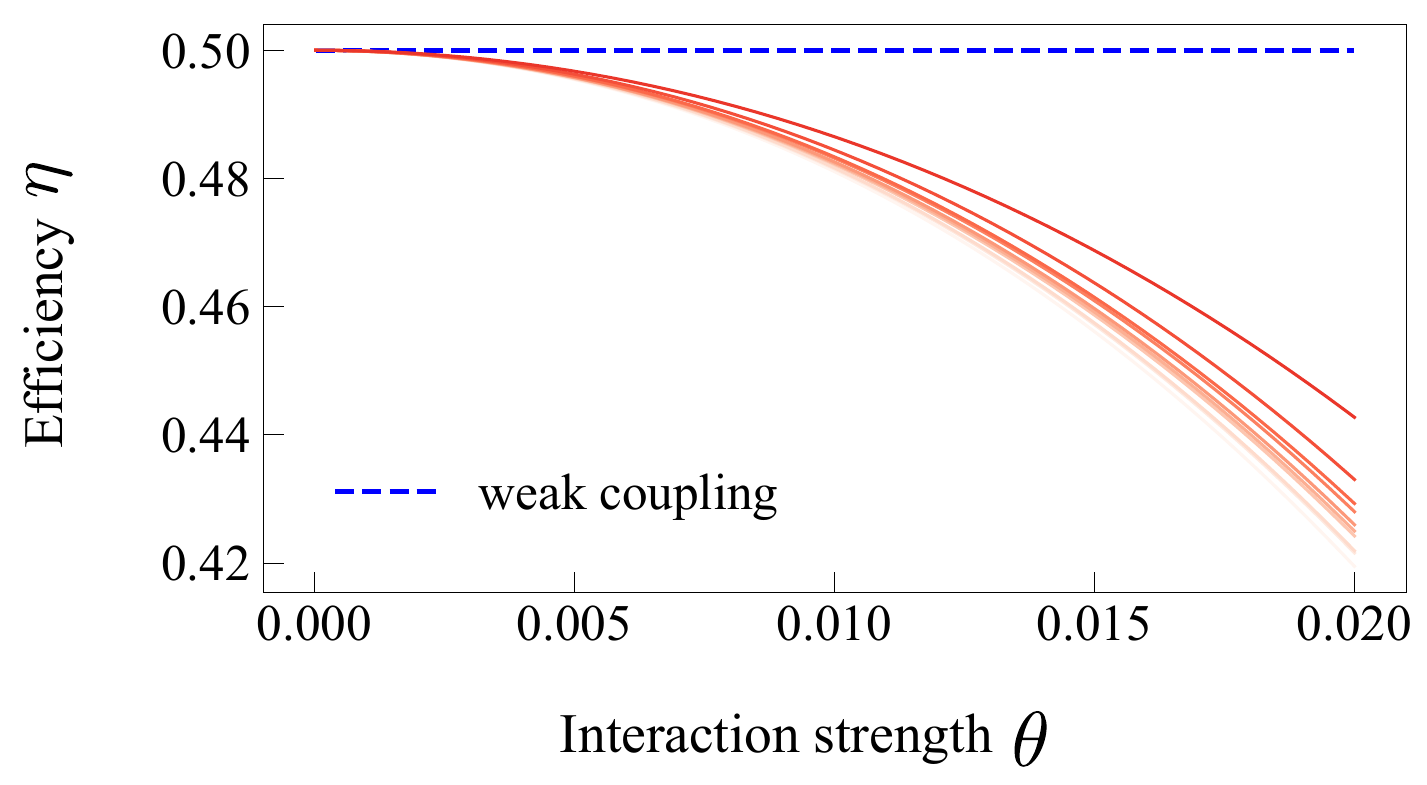}
  \caption{Relationship between interaction strength $\theta$ and efficiency for the two quantum Otto cycle models. The interaction Hamiltonians of the strong coupling model are generated repeatedly and the efficiency for each is plotted. The parameters are set as follows: $\beta_c = 2.0$, $\beta_h = 0.5$, $E_g^c = 0.6$, $E_e^c = 1.4$, $E_g^h = 2E_g^c$, $E_e^h = 2E_e^c$, $\{\epsilon_i^c\}_i \in \{ 0.5, 1.5, ..., 15.5 \}$, $\{ \epsilon_i^h \}_i \in \{ 2.0, 3.0, ..., 17.0 \}$. $\theta$ is varied from 0 to 0.02. The blue dashed line represents $\eta_\mathrm{weak} = 0.5$, the efficiency of the weak coupling model. The red lines show the transition of the efficiency of the strong coupling model with a change in $\theta$. The efficiencies of both models for small $\theta$ are close. As $\theta$ increases, $\eta_\mathrm{str}$ decreases.
  \label{fig:random_sampling}
  }
\end{figure}

We conduct this operation repeatedly.
Figure \ref{fig:random_sampling} shows each of the efficiencies as functions of $\theta$, together with $\eta_\mathrm{weak}$, which can be calculated analytically (see Appendix D).
The blue dashed line represents $\eta_\mathrm{weak}$, which is constant for all $\theta$ because $\eta_\mathrm{weak}$ is calculated neglecting the interaction.
The red lines represent $\eta_\mathrm{str}$.
In the range where $\theta$ is small, $\eta_\mathrm{weak}$ and $\eta_\mathrm{str}$ are very close.
This is consistent with the fact that the strong coupling model agrees with the weak coupling model in the limit of weak interaction.
As the interaction becomes stronger, the efficiency of the strong coupling model decreases.
Although the inequality $\eta_\mathrm{str} \leq\eta_\mathrm{weak}$ has been shown to be violated, our numerical calculation empirically shows that this inequality holds in most cases in the prepared two-level system model.

Finally, we demonstrate the efficiency of the strong coupling model exceeds that of the weak coupling model under specific interaction Hamiltonians.
We prepare the exactly same system and reservoirs as used in the above experiment and fluctuate the eignevalues of the hot reservoir.
Using these quantum states， we generate $U_d^c$ and $U_d^h$ described in Eq. \eqref{Udh} and construct interaction Hamiltonians by Eq. \eqref{Hsb}, which are optimal to minimize the value of $d$ calculated in Eq.\eqref{d_value} and lead to the reversal of the efficiency $\eta_\mathrm{str} > \eta_\mathrm{weak}$.
Then, the efficiency of this model $\eta_\mathrm{str}$ can be calculated numerically.

\begin{figure}
  \includegraphics[width=8.5cm]{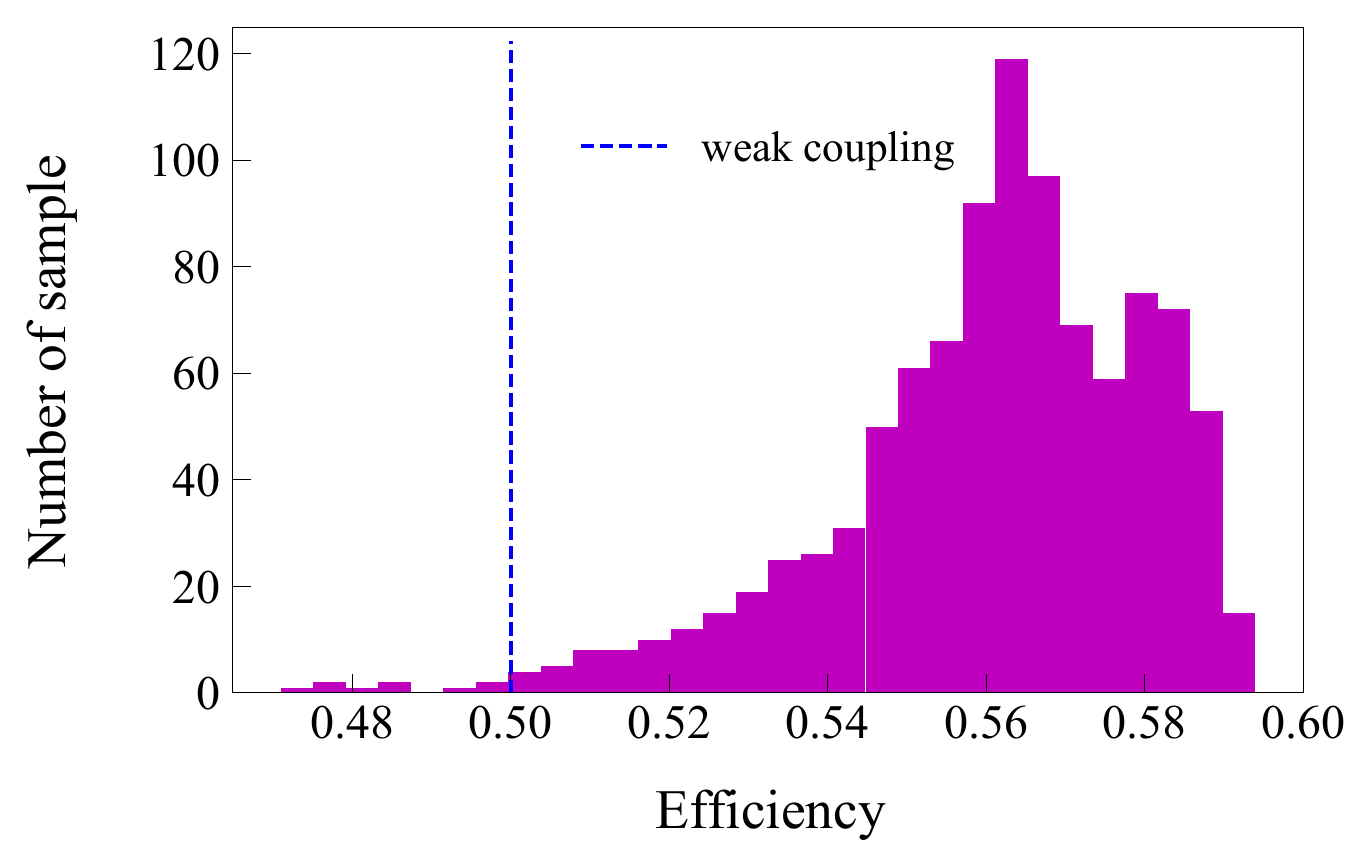}
  \caption{
Distribution of the efficiency of the strong coupling models, which are generated 1000 times under the interaction Hamiltonian designed from Eqs. \eqref{Hsb} and \eqref{Udh}.
  The parameters are set as in Figure \ref{fig:random_sampling}.
  However, each eigenvalue of the hot reservoir is fluctuated by a uniform random value in $[-0.3, 0.3]$.
  The blue dashed line represents $\eta_\mathrm{weak} = 0.5$, the efficiency of the weak coupling model.
  Most values of $\eta_\mathrm{str}$ are higher than $\eta_\mathrm{weak}$.
  \label{fig:reversal}
}
\end{figure}

We repeat this operation and collect the value of $\eta_\mathrm{str}$.
In Figure \ref{fig:reversal}, the pink histogram shows the distribution of $\eta_\mathrm{str}$ and the blue dashed line represents the efficiency of the weak coupling model $\eta_\mathrm{weak} = 0.5$.
We can see the efficiency of the strong coupling model $\eta_\mathrm{str}$ exceeds the value of $\eta_\mathrm{weak}$ in most cases.
This is consistent with the fact that the designed interaction Hamiltonian using Eqs. \eqref{Hsb} and \eqref{Udh} can realize the reversal of the efficiency $\eta_\mathrm{str} > \eta_\mathrm{weak}$.
Yet, the reversal does not occur in rare cases despite the use of the designed interaction Hamiltonian.
This is because the contribution of the term $D(\tilde{\rho}_B^h || \rho_B^h)$ was neglected when we defined $d$ in Eq. \eqref{d_value} to compare the efficiency of the two model.

\section{Conclusion}
In this study, we developed a quantum Otto cycle model, considering the interaction between the system and the reservoirs, which is neglected in the existing weak coupling model.
We emphasize that the proposed model does not specify the details of the system and the reservoirs, and thus, it can be generally applied.
We confirmed that our strong coupling model agrees with the weak coupling model in the limit of weak interaction.
The efficiencies of both models were expressed in closed form and we derived the sufficient condition for $\eta_\mathrm{str} \leq \eta_\mathrm{weak}$.
This condition is satisfied when the decoupling processes introduced to the strong coupling model have positive costs.
Using numeric calculations, we demonstrated that $\eta_\mathrm{str} \simeq \eta_\mathrm{weak}$ in the range of weak interaction, and that when the interaction is not negligible, $\eta_\mathrm{str}$ is lower than $\eta_\mathrm{weak}$ in the prepared system.
These results indicate that our strong coupling model can be regarded as a reasonable extension of the existing weak coupling model.
Furthermore, by more detailed analysis, we suggested the possibility of the reversal of the efficiency $\eta_\mathrm{str} > \eta_\mathrm{weak}$.
We presented a design method to realize the optimal interaction Hamiltonians, which are expected to provide the maximum efficiency of the strong coupling model.
We also confirmed numerically that the strong coupling model achieved higher efficiency compared with the weak coupling model under the interaction Hamiltonians generated by the proposed method.
These are interesting and unique results for our treatment.

We hope that the present work will contribute to the development of a strong-coupling thermodynamics theory.
Additionally, our study is expected to enhance the understanding of quantum correlation because incorporating the interaction terms is synonymous with considering the quantum correlation between the system and the reservoirs.
Thus, we anticipate that this study contributes to several quantum thermodynamics fields, not just quantum heat engines.

\begin{acknowledgments}
This work was supported by KAKENHI Grant Numbers JP19K12153 and JP22H03659.
\end{acknowledgments}

\begin{widetext}

\appendix

\section{Heat transferred in thermalization process}
The heat transferred in the thermalization processes can be expressed by the von Neumann entropy and the quantum relative entropy.
Here, we derive the expressions in both the weak coupling model and the strong coupling model.

In the weak coupling model, the quantum relative entropy $D(\rho_S^h || \rho_S^{h, \mathrm{eq}})$ can be calculated as
\begin{align}
  D(\rho_S^h || \rho_S^{h, \mathrm{eq}}) &= \mathrm{Tr}[\rho_S^h\ln\rho_S^h] - \mathrm{Tr}[\rho_S^h\ln\rho_S^{h, \mathrm{eq}}] \notag \\
  &= S(\rho_S^{h, \mathrm{eq}}) - S(\rho_S^h) + \mathrm{Tr}[\rho_S^{h, \mathrm{eq}}\ln\rho_S^{h, \mathrm{eq}}] - \mathrm{Tr}[\rho_S^h\ln\rho_S^{h, \mathrm{eq}}] \notag \\
  &= \Delta S + \mathrm{Tr}\left[ (\rho_S^{h,\mathrm{eq}} - \rho_S^h)(-\beta_hH_S^h - \ln Z_S^h) \right] \notag \\
  &= \Delta S -\beta_h\left(\mathrm{Tr}[H_S^h\rho_S^{h, \mathrm{eq}}] - \mathrm{Tr}[H_S^h\rho_S^h]\right) \notag \\
  &= \Delta S -\beta_hQ_{\mathrm{in}},
\end{align}
where $\Delta S = S(\rho_S^{h,\mathrm{eq}}) - S(\rho_S^h)$.
From this relation, we obtain Eq.~\eqref{q_in}.
Furthermore, the following relation holds:
\begin{equation}
  \Delta S = \beta_hQ_\mathrm{in} + D(\rho_S^h||\rho_S^{h,\mathrm{eq}}).
\end{equation}
This equality shows that the change in the von Neumann entropy of the system in the thermalization process can be divided into two contributions: the heat flux and the entropy production \cite{esposito2010entropy}.
Similarly, we can derive Eq.~\eqref{q_out} from the following relation:
\begin{align}
  D(\rho_S^c||\rho_S^{c,\mathrm{eq}}) &= S(\rho_S^{c,\mathrm{eq}}) - S(\rho_S^c) -\beta_c\left( \mathrm{Tr}[H_S^c\rho_S^{c,\mathrm{eq}}] - \mathrm{Tr}[H_S^c\rho_S^c] \right) \notag \\
  &= -\Delta S + \beta_cQ_\mathrm{out}.
\end{align}

Next, we describe the heat transferred in the thermalization processes of the strong coupling model.
$D(\rho_{SB}^h||\tilde{\rho}_S^h\otimes\rho_B^h)$ and $D(\rho_S^h\otimes\rho_B^h||\rho_{SB}^h)$ can be calculated as follows:
\begin{align}
  D(\rho_{SB}^h||\tilde{\rho}_S^h\otimes\rho_B^h) &= \mathrm{Tr}[\rho_{SB}^h\ln\rho_{SB}^h] - \mathrm{Tr}[\tilde{\rho}_S^h\ln\tilde{\rho}_S^h] - \mathrm{Tr}[\tilde{\rho}_B^h\ln\rho_B^h] \notag \\
  &= -S(\rho_{SB}^h) + S(\tilde{\rho}_S^h) + S(\rho_B^h) + \mathrm{Tr}[(\rho_B^h - \tilde{\rho}_B^h)\ln\rho_B^h] \notag \\
&= -S(\rho_{SB}^h) + S(\tilde{\rho}_S^h) + S(\rho_B^h) -\beta_h\mathrm{Tr}[H_B^h(\rho_B^h - \tilde{\rho}_B^h)],
\end{align}
\begin{align}
  D(\rho_S^h\otimes\rho_B^h||\rho_{SB}^h) &= -S(\rho_S^h\otimes\rho_B^h) - \mathrm{Tr}[(\rho_S^h\otimes\rho_B^h)\ln\rho_{SB}^h] \notag \\
  &= -S(\rho_S^h) -S(\rho_B^h) + S(\rho_{SB}^h) + \mathrm{Tr}[(\rho_{SB}^h - \rho_S^h\otimes\rho_B^h)\ln\rho_{SB}^h] \notag \\
  &= S(\rho_{SB}^h) - S(\rho_S^h) - S(\rho_B^h) + \mathrm{Tr} \left[(\rho_{SB}^h - \rho_S^h\otimes\rho_B^h) \left\{ -\beta_h(H_S^h+H_B^h+H_{SB}^h)-\ln Z_{SB}^h \right\} \right] \notag \\
  &= S(\rho_{SB}^h) - S(\rho_S^h) - S(\rho_B^h) -\beta_h \left\{ \mathrm{Tr}[H_S^h(\tilde{\rho}_S^h-\rho_S^h)] + \mathrm{Tr}[H_B^h(\tilde{\rho}_B^h-\rho_B^h)] + \mathrm{Tr}[H_{SB}^h\rho_{SB}^h] \right\}.
\end{align}
From these two equalities, we obtain 
\begin{equation}
  D(\rho_{SB}^h||\tilde{\rho}_S^h\otimes\rho_B^h) + D(\rho_S^h\otimes\rho_B^h||\rho_{SB}^h) = S(\tilde{\rho}_S^h) - S(\rho_S^h) -\beta_h \left\{ \mathrm{Tr}[H_S^h(\tilde{\rho}_S^h-\rho_S^h)] + \mathrm{Tr}[H_{SB}^h\rho_{SB}^h] \right\},
\end{equation}
and therefore, Eq.~\eqref{q_th_h} holds.
By the same calculations for the low temperature side, we can derive Eq.~\eqref{q_th_c}.

\section{Heat transferred in decoupling process}
In this section, we introduce the definition of heat proposed in Ref.~\cite{xu2018achieving} and calculate the heat transferred in the decoupling processes of the strong coupling model.
We discuss Process B-2 and calculate only $Q_d^h$.
However, $Q_d^c$ can be calculated in the same way.

$\rho$ represents the state of the compound system $S+B_h$ in Process B-2.
$\rho_S$ and $\rho_B$ are the reduced states: $\rho_S = \mathrm{Tr}_B[\rho]$ and $\rho_B = \mathrm{Tr}_S[\rho]$, respectively.
We define the heat transferred from the reservoir to the system during an infinitesimal time $dt$ as $dQ = -i\mathrm{Tr}\left[ [H_S^\mathrm{eff}, H_\mathrm{tot}]C \right]dt$.
Here, $H_S^\mathrm{eff}$ is the effective Hamiltonian of the system, defined as $H_S^\mathrm{eff} = H_\mathrm{tot} - H_B^\mathrm{eff}$, where $H_B^\mathrm{eff} = -\frac{1}{\beta_h}\ln\rho_B$.
$H_\mathrm{tot}$ is the Hamiltonian of the compound system, including the interaction.
$C=\rho-\rho_S\otimes\rho_B$ corresponds to the quantum coherence of $\rho$.
Using these definitions, $dQ$ can be rewritten as follows:
\begin{align}
  dQ &= -i\mathrm{Tr}\left[ [H_S^\mathrm{eff}, H_\mathrm{tot}]C \right]dt \notag \\
  &= -i\mathrm{Tr}\left[ [-H_B^\mathrm{eff}, H_\mathrm{tot}]C \right]dt \notag \\
  &= -i\frac{1}{\beta_h}\mathrm{Tr}\left[ [\mathbb{I}_S\otimes\ln\rho_B, H_\mathrm{tot}]C \right]dt.
\end{align}
On the other hand, the infinitesimal change in von Neumann entropy of $\rho_S$ can be calculated as follows:
\begin{align}
  dS_S &= -d \mathrm{Tr}[\rho_S\ln\rho_S] \notag \\
  &= -d \mathrm{Tr}[\rho(\ln\rho_S\otimes\mathbb{I}_B)] \notag \\
  &= -\mathrm{Tr}[(\ln\rho_S\otimes\mathbb{I}_B)d\rho] \notag \\
  &= i \mathrm{Tr}\left[ [H_\mathrm{tot}, \rho](\ln\rho_S\otimes\mathbb{I}_B) \right]dt \notag \\
  &= i \mathrm{Tr}\left[ [\ln\rho_S\otimes\mathbb{I}_B, H_\mathrm{tot}]\rho \right]dt \notag \\
  &= i \mathrm{Tr}\left[ [\ln\rho_S\otimes\mathbb{I}_B, H_\mathrm{tot}]C \right]dt.
\end{align}
We used the von Neumann equation $d\rho = -i[H_\mathrm{tot}, \rho]dt$ for the fourth line above and $[\ln\rho_S\otimes\mathbb{I}_B, \rho_S\otimes\rho_B]=0$ for the last line.
Therefore, we can derive the following relation:
\begin{align}
  \label{b3}
  dS_S -\beta_hdQ &= i \mathrm{Tr}\left[ [\ln(\rho_S\otimes\rho_B), H_\mathrm{tot}] C \right] dt \notag \\
  &= i \mathrm{Tr}\left[ [\ln(\rho_S\otimes\rho_B), H_\mathrm{tot}] \rho \right] dt \notag \\
  &= i \mathrm{Tr}\left[ \ln(\rho_S\otimes\rho_B)[H_\mathrm{tot}, \rho] \right] dt \notag \\
  &= - \mathrm{Tr}[\ln(\rho_S\otimes\rho_B) d\rho] \notag \\
  &= -d \mathrm{Tr}[\rho\ln(\rho_S\otimes\rho_B)].
\end{align}
By integrating both sides of Eq.~\eqref{b3} from the initial state $\rho_{SB}^h$ to the final state $\rho_S^{h,\mathrm{eq}}\otimes\rho_B^h$, the following relation is obtained:
\begin{align}
  S(\rho_S^{h,\mathrm{eq}}) - S(\tilde{\rho}_S^h) - \beta_hQ_d^h &= -\mathrm{Tr}[(\rho_S^{h,\mathrm{eq}}\otimes\rho_B^h)\ln(\rho_S^{h,\mathrm{eq}}\otimes\rho_B^h)] + \mathrm{Tr}[\rho_{SB}^h\ln(\tilde{\rho}_S^h\otimes\tilde{\rho}_B^h)] \notag \\
  &= - \mathrm{Tr}[\rho_{SB}^h\ln\rho_{SB}^h] + \mathrm{Tr}[\rho_{SB}^h\ln(\tilde{\rho}_S^h\otimes\tilde{\rho}_B^h)] \notag \\
  &= -D(\rho_{SB}^h||\tilde{\rho}_S^h\otimes\tilde{\rho}_B^h) \notag \\
\beta_hQ_d^h &= S(\rho_S^{h,\mathrm{eq}}) - S(\tilde{\rho}_S^h) + D(\rho_{SB}^h||\tilde{\rho}_S^h\otimes\tilde{\rho}_B^h).
\end{align}
This is equal to Eq.~\eqref{q_d_h}.
Note that we used the restriction $S(\rho_S^{h,\mathrm{eq}}\otimes\rho_B^h) = S(\rho_{SB}^h)$ for the second line in the relation above.
By the same calculations, we can derive Eq.~\eqref{q_d_c}.

\section{Majorization relation}
Here, we review the general properties of majorization relation and provide the rigorous proof for Eq. \eqref{majo_ineq} in the main text.

Let $x,y \in \mathbb{R}^n$.
$x^\downarrow \coloneqq (x_1^\downarrow, ..., x_n^\downarrow)^\top$ and $y^\downarrow \coloneqq (y_1^\downarrow, ..., y_n^\downarrow)^\top$ are defined as the vectors of sorted elements of $x,y$ such that $x_1^\downarrow \geq ... \geq x_n^\downarrow$ and $y_1^\downarrow \geq ... \geq y_n^\downarrow$.
We say $x$ majorizes $y$ when the following two conditions are satisfied:
\begin{align}
    \forall k \in \{1,2,...,n\},& \;\; \sum_{i=1}^ky_i^\downarrow \leq \sum_{i=1}^k x_i^\downarrow, \\
    \sum_{i=1}^n x_i &= \sum_{i=1}^n y_i,
\end{align}
and this relation is written by $y \prec x$.
Moreover, if a square matrix $A = (a_{ij})$ satisfies $a_{ij}\geq 0$ and $\sum_i a_{ij} = \sum_j a_{ij} = 1$, $A$ is a doubly stochastic matrix.
In general, $n\times n$ doubly stochastic matrix $A$ holds $Ax \prec x$ for any $n$-dimensional real vector \cite{marshall1979inequalities,sagawa2022entropy}.

Next is the proof for the following proposition: for any $n$-dimensional real vectors $w,x,y$ such that $w_1\leq .. \leq w_n$, $x_1\geq ... \geq x_n$ and $y_1\geq ... \geq y_n$, if $y\prec x$, then $w^\top y \geq w^\top x$.
First, when $n=2$, this inequality is satisfied: from $x_1-y_1 = y_2-x_2 \geq 0$, $w^\top y - w^\top x = w_1(y_1-x_1) + w_2(y_2-x_2) = (w_2-w_1)(x_1-y_1) \geq 0$.
Next, we assume that for some $n$ the inequality holds.
Then, for $n+1$-dimensional vectors $w,x,y$, the same inequality is satisfied:
\begin{align}
\label{majo_proof}
    w^\top y - w^\top x &= \sum_{i=1}^{n+1}w_i(y_i-x_i) \notag \\
    &= \sum_{i=1}^n w_i(y_i-x_i) + w_{n+1}(y_{n+1} - x_{n+1}) \notag \\
    &\geq \sum_{i=1}^n w_i(y_i-x_i) + w_n(y_{n+1}-x_{n+1}) \notag \\
    &= w'^\top y' - w'^\top x' \notag \\
    &\geq 0.
\end{align}
Here, $w' = (w_1, ..., w_n)^\top$, $x'=(x_1, ..., x_n)^\top$ and $y'=(y_1, ..., y_{n-1}, y_n+y_{n+1}-x_{n+1})$ are $n$-dimensional real vectors.
We note that when $y\prec x$, $y'\prec x'$ is also satisfied.
From the supposition of mathematical induction, the last inequality of Eq. \eqref{majo_proof} holds.
Therefore, for any $n$, $w^\top y \geq w^\top x$.

Using this proposition, we can prove the Eq. \eqref{majo_ineq} briefly.
Recall that $\boldsymbol{E}$ and $\boldsymbol{P}_\downarrow$ are real vectors whose elements are ordered ascending and descending order, respectively.
$A$ is a doubly stochastic matrix.
Eq. \eqref{majo_ineq} is shown as below:

\begin{equation}
\label{majo_concl}
    \boldsymbol{E^\top AP} \geq \boldsymbol{E^\top (AP)_\downarrow} \geq \boldsymbol{E^\top P_\downarrow},
\end{equation}
where $(\boldsymbol{AP})_\downarrow$ is a vector of $\boldsymbol{AP}$ in descending order.
The last inequality of Eq. \eqref{majo_concl} is from the proposition proved above, because $(\boldsymbol{AP})_\downarrow \prec \boldsymbol{P}_\downarrow$ is followed immediately from $\boldsymbol{AP} \prec \boldsymbol{P}$.

\section{Analytical calculation of efficiency}
This section fully describes the calculation of the efficiency of the weak coupling model prepared for the simulation.
We signify the probability distributions of $\rho_S^{c,\mathrm{eq}}$ and $\rho_S^{h,\mathrm{eq}}$ as follows:
\begin{equation}
  p_g^c = \frac{e^{-\beta_c E_g^c}}{Z_S^c},\;\;\; p_e^c = \frac{e^{-\beta_c E_e^c}}{Z_S^c},\;\;\; p_g^h = \frac{e^{-\beta_h E_g^h}}{Z_S^h},\;\;\; p_e^h = \frac{e^{-\beta_h E_e^h}}{Z_S^h}.
\end{equation}
Using these values, the von Neumann entropy and the quantum relative entropy can be calculated as
\begin{align}
  \Delta S - D(\rho_S^h||\rho_S^{h,\mathrm{eq}}) &= S(\rho_S^{h,\mathrm{eq}}) - S(\rho_S^h) + S(\rho_S^h) + \mathrm{Tr}[\rho_S^h\ln\rho_S^{h,\mathrm{eq}}] \notag \\
  &= (p_g^c - p_g^h)\ln p_g^h + (p_e^c - p_e^h)\ln p_e^h \notag \\
  &= (p_g^c - p_g^h)(-\beta_hE_g^h -\ln Z_S^h) + (p_e^c - p_e^h)(-\beta_hE_e^h -\ln Z_S^h) \notag \\
  &= \beta_h\left\{ (p_g^h-p_g^c)E_g^h + (p_e^h-p_e^c)E_e^h \right\},
\end{align}
\begin{align}
  \Delta S + D(\rho_S^c||\rho_S^{c,\mathrm{eq}}) &= S(\rho_S^c) - S(\rho_S^{c,\mathrm{eq}}) - S(\rho_S^c) - \mathrm{Tr}[\rho_S^c\ln\rho_S^{c,\mathrm{eq}}] \notag \\
  &= (p_g^c - p_g^h)\ln p_g^c + (p_e^c - p_e^h)\ln p_e^c \notag \\
  &= (p_g^c - p_g^h)(-\beta_cE_g^c -\ln Z_S^c) + (p_e^c - p_e^h)(-\beta_cE_e^c -\ln Z_S^c) \notag \\
  &= \beta_c\left\{ (p_g^h-p_g^c)E_g^c + (p_e^h-p_e^c)E_e^c \right\}.
\end{align}
We used $p_g^c + p_e^c = p_g^h + p_e^h = 1$ for the last line of both calculations.
Thus, from Eq.~\eqref{eta_weak}, we can rewrite $\eta_\mathrm{weak}$ as
\begin{equation}
  \eta_\mathrm{weak} = 1 - \frac{(p_g^h-p_g^c)E_g^c + (p_e^h-p_e^c)E_e^c}{(p_g^h-p_g^c)E_g^h + (p_e^h-p_e^c)E_e^h}.
\end{equation}
By setting $E_g^h = 2E_g^c$ and $E_e^h = 2E_e^c$, we obtain
\begin{equation}
  \eta_\mathrm{weak} = 1 - \frac{1}{2} = \frac{1}{2}.
\end{equation}

\end{widetext}

\end{document}